\documentclass[conference,usenames,dvipsnames,x11names,table]{IEEEtran}
\usepackage[abbreviations]{foreign}
\usepackage[binary-units]{siunitx}
\usepackage[utf8]{inputenc}
\usepackage{graphicx}
\usepackage{xcolor}  
\usepackage{amsmath,amssymb}
\usepackage{algorithm,}
\usepackage[noend]{algpseudocode}
\usepackage{ifthen}
\usepackage{multicol}
\usepackage{enumitem}
\usepackage{color}
\usepackage{listings}
\usepackage{pifont}
\usepackage{multirow}
\usepackage[hyphens,spaces,obeyspaces]{url}
\usepackage{hyperref}
\usepackage{cleveref} % load after hyperref
\usepackage{amsthm}
\usepackage{array}% http://ctan.org/pkg/array
\usepackage{float}
\usepackage{booktabs}
\usepackage{subcaption}
\usepackage[backend=bibtex,sorting=none,style=ieee,url=false,isbn=false]{biblatex}

\newboolean{showcomments}
\setboolean{showcomments}{false}
\ifthenelse{\boolean{showcomments}}
{ \newcommand{\mynote}[3]{
   \fbox{\bfseries\sffamily\scriptsize#1}
   {\small$\blacktriangleright$\textsf{\emph{\color{#3}{#2}}}$\blacktriangleleft$}}}
{ \newcommand{\mynote}[3]{}}

\newcommand{\system}{\textsc{Heats}\xspace}
\newcommand{\sys}{\system}

\newcommand{\cm}{{\color{Green4}\ding{51}}}% check mark
\newcommand{\xm}{{\color{BrickRed}\ding{55}}}% x mark

\bibliography{paper}

\renewbibmacro{finentry}{%
    \iffieldequalstr{entrykey}{brenner2016securekeeper}%<- key after which you want the break
    {\finentry\newpage}
    {\finentry}}

\begin{document}
\title{\sys: Heterogeneity- and Energy-Aware\\Task-based Scheduling}

\author{
\\[-5.0ex]

\IEEEauthorblockN{
Isabelly Rocha\IEEEauthorrefmark{1},
Christian G\"ottel\IEEEauthorrefmark{1},
Pascal Felber\IEEEauthorrefmark{1},
Marcelo Pasin\IEEEauthorrefmark{1},
Romain Rouvoy\IEEEauthorrefmark{2} and
Valerio Schiavoni\IEEEauthorrefmark{1}
}

\IEEEauthorblockA{
\IEEEauthorrefmark{1}University of Neuch\^atel, Switzerland, first.last@unine.ch}

\IEEEauthorblockA{
\IEEEauthorrefmark{2}Inria Lille – Nord Europe, first.last@inria.fr}

\\[-5.0ex]
}

\maketitle
% !TEX root = paper.tex
\begin{abstract}
    %\vs{MAX 500 words, 10 pages ALL included}
Cloud providers usually offer diverse types of hardware for their users.
Customers exploit this option to deploy cloud instances featuring GPUs, FPGAs, architectures other than x86 (\eg, ARM, IBM Power8), or featuring certain specific extensions (\eg, Intel SGX).
We consider in this work the instances used by customers to deploy containers, nowadays the \emph{de facto} standard for micro-services, or to execute computing tasks.
In doing so, the underlying container orchestrator (\eg, Kubernetes) should be designed so as to take into account and exploit this hardware diversity.
In addition, besides the feature range provided by different machines, there is an often overlooked diversity in the energy requirements introduced by hardware heterogeneity, which is simply ignored by default container orchestrator's placement strategies.
%Moreover, orchestration systems rely on scheduling policies that know beforehand the lifespan and resource requirements of containerized applications.
We introduce \sys, a new task-oriented and energy-aware orchestrator for containerized applications targeting heterogeneous clusters.
\sys allows customers to trade performance vs. energy requirements.
Our system first learns the performance and energy features of the physical hosts.
Then, it monitors the execution of tasks on the hosts and opportunistically migrates them onto different cluster nodes to match the customer-required deployment trade-offs. 
%These predictions are used to expose clients to different deployment trade-offs (\ie performance, energy savings).
%\vs{(How) should do we tackle locality?}
%We use GenPack, a framework to schedule containers in a cloud data center leveraging principles from generational garbage collection (GC). 
%By enhancing its runtime monitoring system, we can estimate the lifetime as well as the energy requirements and schedule the containers on appropriate cloud resources. 
Our \sys prototype is implemented within Google's Kubernetes. 
The evaluation with synthetic traces in our cluster indicate that our approach can yield considerable energy savings (up to 8.5\%) and only marginally affect the overall runtime of deployed tasks (by at most 7\%).
\sys is released as~open-source.

%Our heterogeneity- and energy-aware scheduler is compared against the default Kubernetes scheduler by means of real-life traces.
\end{abstract}

% !TEX root = paper.tex
\section{Introduction}
\vspace{-2pt}

%\vs{what follows is a possible sketch of story line to explain: 1) what we want to achieve, 2) why this is important, 3) why other systems (as there seems to be lots of RW) do not provide this out of the box or cannot be modified to provide this (this implies looking listing and understanding how those systems work), and 4) what are the technical, system challenges to solve toward 1). Don't take this for granted yet.}

Cloud providers nowadays provide access to a wide range of heterogeneous resources to their customers.
Hence, the diversity of resources encourages application developers and deployers to program for, and offload even more workloads to, the cloud.
There, specialized hardware (\eg, GPU, FPGA) can be rented for limited time, reducing upfront costs and allowing for better scalability. 

To illustrate this diversity, Table~\ref{tab:cloudresources} shows an overview of the commercial offering of heterogeneous resources at 6 major public cloud providers.
For each, we list the CPU architecture (x86, IBM Power, ARM), and the availability of GPU, FPGA or ASIC units. 
We further indicate if such resources can be accessed using bare metal (BM) or virtual machine (VM) instances.
Additionally, we show whether the operating frequency of the processor can be dynamically scaled up or down, a feature that could be leveraged to reduce the generated energy costs of a node.
This quick survey reveals that it is possible to combine a very heterogeneous ensemble of machines, each offering specific hardware feature sets.
This capability represents the ideal case for applications that have different resource demands, as it is sometimes better to migrate the execution from a machine of one kind to a different one, in order to better match the expected trade-off requested by the customer.
Resource diversity can also be exploited to deploy applications and workloads of different nature.

\begin{figure}[!t]
	\centering
	\includegraphics[scale=0.7]{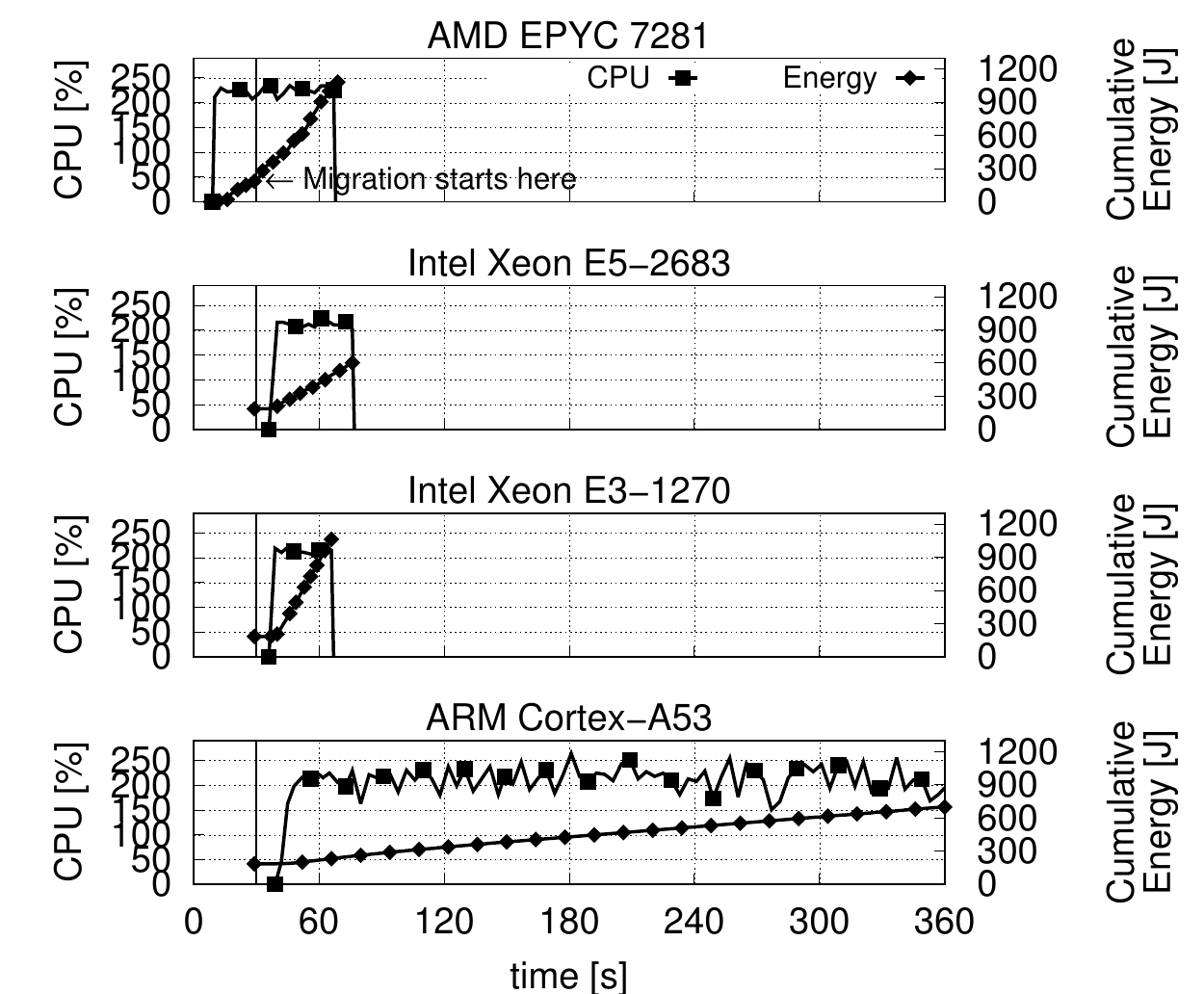}
	\caption{Migrating a task to a different host allows for energy savings but increased run time.}
	\label{fig:motivating_example}
    \vspace{-10pt}
\end{figure}

Containers (\emph{e.g.}, Docker~\cite{merkel2014docker}) have recently become the de facto standard to deploy applications on the cloud, executed by specialized container orchestrators, such as Google's Kubernetes~\cite{medel2016modelling}.
Current policies of container orchestrators often ignore the diversity found in hardware, leading to subtle trade-off between energy and performance.
To better understand this aspect and motivate our work, we conducted a simple experimental study (Figure~\ref{fig:motivating_example}).
We set up an on-premise cluster composed of 4 different types of nodes: three server-grade machines (two Intel and one AMD) and one ARM-based low-energy device (a Raspberry Pi). 
Each machine has different hardware characteristics (\emph{e.g.}, number and type of CPU cores, memory and operating frequency) and energy requirements.
% (we detail the specificities of this cluster in Section~\ref{sec:eval}).

\begin{table}[!t]
  \centering
  \setlength{\aboverulesep}{0pt}
  \setlength{\belowrulesep}{0pt}
  \setlength{\extrarowheight}{.74ex}
  \setlength\tabcolsep{1.5pt}
  \rowcolors{1}{gray!10}{gray!0}
  \vspace{4pt}
  \caption{Heterogenous resources available at public cloud providers. Some types available only via bare metal (BM) or virtual machines (VM).  \textbf{*} = frequency scaling enabled. \cm = feature available from VM or BM. \xm = not available.}
  \begin{tabular}{>{\kern-\tabcolsep}lcccccccccccc<{\kern-\tabcolsep}}
    \toprule \rowcolor{gray!25}
     & \multicolumn{2}{c}{\textbf{x86-64}} & \multicolumn{2}{c}{\textbf{POWER}} & \multicolumn{2}{c}{\textbf{ARM}} & \multicolumn{2}{c}{\textbf{GPU}} & \multicolumn{2}{c}{\textbf{FPGA}} & \multicolumn{2}{c}{\textbf{ASIC}} \\ \rowcolor{gray!25}
    \multirow{-2}{*}{\textbf{Provider}} & \textbf{BM} & \textbf{VM} & \textbf{BM} & \textbf{VM} & \textbf{BM} & \textbf{VM} & \textbf{BM} & \textbf{VM} & \textbf{BM} & \textbf{VM} & \textbf{BM} & \textbf{VM} \\
    \midrule
    Amazon~\cite{aws:iaas}           & \cm* & \cm & \xm & \xm & \xm & \xm & \xm & \cm & \xm & \cm & \xm & \xm \\
    Microsoft~\cite{azure:iaas}       & \xm & \xm & \xm & \xm & \xm & \xm & \xm & \cm & \xm & \cm & \xm & \xm \\
    Google~\cite{gcp:iaas}           & \xm & \cm & \xm & \xm & \xm & \xm & \xm & \cm & \xm & \xm & \xm & \cm \\
    IBM~\cite{ibm:iaas}           & \cm* & \xm & \cm* & \xm & \xm & \xm & \cm & \xm & \xm & \xm & \xm & \xm \\
    Oracle~\cite{oracle:iaas}     & \cm* & \cm* & \xm & \xm & \xm & \xm & \cm & \cm & \xm & \xm & \xm & \xm \\
    Scaleway~\cite{scaleway:iaas} & \cm* & \xm & \xm & \xm & \cm* & \xm & \xm & \xm & \xm & \xm & \xm & \xm \\
    \bottomrule
  \end{tabular}
  \label{tab:cloudresources}
  \vspace{-15pt}
\end{table}

While these properties are known by the cluster owner at deployment time, the energy requirements as well as the raw computing power of the machines for a specific workload are not.
Typically, customers are only able to evaluate those at runtime, while executing their applications. 
Because of that, they can face unexpected costs or missed deadlines upon completion of tasks.

In our scenario, we developed and deployed a simple task implementing the popular \emph{k-means} clustering algorithm.
At first, the task is deployed on the AMD node (Figure~\ref{fig:motivating_example}, top-most plot).
Given our cluster settings, with the default Kubernetes scheduler, we observe the deployment on the machine with more cores and memory.
When remaining in the same host, the task completes after 69 seconds, consuming 1,047 Joules.

Next, we consider customers wishing to compromise the running time for energy costs.
This requires a dynamic container rescheduling policy that can migrate a task into the ARM node after it has made some progress but before completion (e.g., 30 seconds after startup, as highlighted by the vertical line in each plot).
In doing so, the net energy savings are important (up to 34\%) but at the cost of a $5.4\times$ increase of the task's running time.

Such trade-offs are often desirable (especially for deadline-free, low-priority workloads), but difficult to achieve in practice.
As a matter of fact, a task (or container) orchestrator would need to be aware of several factors and able to:
(1)~know or learn the characteristics of the underlying cluster and its hardware resources;
(2)~understand the trade-off that a customer is willing to accept;
(3)~observe if a better placement opportunity exists for the currently executing tasks; and
(4)~migrate the task accordingly.
In this paper we introduce \sys, a scheduling system geared toward heterogeneous clusters that achieves these goals.

The key mechanism used by \sys consists in offering to clients the ability to indicate, at deployment time, their intended energy-performance ratio (the acceptable \emph{trade-off}), in the form of an \emph{H} value.
Thereafter, \sys continuously matches the demanded \emph{H} value to the available resources, considering the resources themselves, pre-built performance and energy models, and the possibly conflicting requirements from other concurrent tasks.
As shown in Section~\ref{sec:eval}, this has consequences on the task throughput of the underlying cluster.

In summary, our \textbf{contributions} are as follows:
\begin{itemize}
    \item We present a probing framework, which we use to build a model of the underlying hardware resources;
    \item We design and implement \sys, a new container scheduler system that, by leveraging the underlying model, places application tasks onto the best matching nodes among the currently available hardware resources for the intended energy/performance ratio;
    \item We thoroughly evaluate our prototype by means of an in-depth experimental evaluation.
\end{itemize}

The rest of this paper is organized as follows.
The rational of the \sys scheduling policy is presented in Section~\ref{sec:heats}.
We describe the architecture of \sys in Section~\ref{sec:architecture}.
We then provide insights on the implementation of the \sys prototype in Section~\ref{sec:implementation}.
%We describe in detail the synthetic \vs{and real-world} traces used in our evaluation in Section~\ref{sec:traces}.
We extensively evaluate the performance of our prototype in Section~\ref{sec:eval}, where we also detail the synthetic %\vs{and real-world} 
traces used to show the benefits of \sys.
We survey related work in Section~\ref{sec:rw}, before concluding in Section~\ref{sec:conclusion}.

\section{\sys Scheduling Policy}
\label{sec:heats}
\vspace{-2pt}
In this section we describe the scheduling algorithm implemented by \sys.
%As mentioned earlier, upon the deployment of a new task, end-users specify their desired power/energy trade-off level.
Algorithm~\ref{alg:heats} describes the main functions, which we detail next.

The resource requirements of a task, as for instance memory or number of cores, are specified before submission.
Resource availability in the hardware nodes is monitored (in our practical experiment we used Heapster~\cite{k8s_heapster}) and reported to \sys monitoring module.
Then, \sys computes suitable nodes for execution considering the resource requirements for all previously running tasks as well as the availability reported by the underlying system.
Next, the algorithm executes a profiling phase and estimates the performance and energy requirements of the given task in each of the previously computed available nodes.
Finally, the scheduling module relies on these estimations to compute scores for each node, to be weighted by the energy/performance ratio defined by the client ($e_w$ and $p_w$ in Algorithm~\ref{alg:heats}).  
The best fitting node is chosen to deploy the given task.

\begin{algorithm}[!t]
{%\fontsize{6}{8}
\begin{algorithmic}[1]
\Function{Schedule}{} \vspace{-0.1em}
\While{$\mathit{pendingTasks \neq \varnothing}$}   \vspace{-0.1em}
	\State $\mathit{t = pendingTasks.poll()}$  \vspace{-0.1em} 
	\State $\mathit{bestFit \leftarrow \textsc{BestFit}(t, e_w, p_w)}$   \vspace{-0.1em}
	\State $\mathit{\textsc{Assign}(t, bestFit)}$   \vspace{-0.1em}
\EndWhile \vspace{-0.1em}
\EndFunction
\Function{Reschedule}{}  \vspace{-0.1em}
\For{$\mathit{t \in runningTasks}$} \vspace{-0.1em}
	\State $\mathit{bestFit \leftarrow \textsc{BestFit}(t, e_w, p_w)}$ \vspace{-0.1em}
	\If{$\mathit{bestFit \neq currentHost}$} \vspace{-0.1em}
		\State $\mathit{\textsc{Migrate}(t, currentHost, bestFit)}$ \vspace{-0.1em}
	\EndIf \vspace{-0.1em}
\EndFor
\EndFunction
\Function{BestFit}{$\mathit{t, e_w, p_w}$} \vspace{-0.1em}
	\State $\mathit{r \leftarrow \textsc{RequiredResources}(t)}$ \vspace{-0.1em}
	\State $\mathit{n \leftarrow \textsc{AvailableNodes}(r)}$ \vspace{-0.1em}
	\State $\mathit{scores \leftarrow \textsc{Scores}(n, r, e_w, p_w)}$ \vspace{-0.1em}
	\State \Return $\mathit{n \; | \; \{n,s\} \in scores \wedge s = max_s}$ \vspace{-0.1em}
\EndFunction \vspace{-0.1em}
\Function{Scores}{$\mathit{nodes, r, e_w, p_w}$} \vspace{-0.1em}
	\State $\mathit{scores \leftarrow \varnothing}$ \vspace{-0.1em}
	\State $\mathit{n_e, n_p \leftarrow \textsc{Predict}(nodes, r)}$ \vspace{-0.1em}
	\For{$\mathit{n \in nodes}$} \vspace{-0.1em}
		\State $\mathit{n_s \leftarrow e_w(1-n_e/max_e) + p_w(n_p/max_p)}$ \vspace{-0.1em}
		\State $\mathit{scores.add(\{n,n_s\})}$ \vspace{-0.1em}
	\EndFor \vspace{-0.1em}
	\State \Return $\mathit{scores}$ \vspace{-0.1em}
\EndFunction 
\end{algorithmic}
\caption{Task scheduling in \sys.}\label{alg:heats}
}
\end{algorithm}

In summary, the \sys strategy will attempt to place tasks on the most efficient host that still has enough resources to run the given task.
We define \emph{most efficient} as the closest match to the demanded energy/performance trade-off.
However, the ideal node for a task will not always be available at scheduling time. 
Therefore, we recompute our scheduling decision every now and then.
When a better fit than the current host of a task is found, the scheduler performs a migration.

%Algorithm~\ref{alg:heats} shows the scheduling policy in more details. 
The scheduling phase is triggered for the queue of all pending tasks.
The algorithm starts by finding the best fit for the next task (lines~4 and~11--15). 
It identifies its resource requirements, \eg, CPU and memory, as well as the available nodes for these resources (lines~12--13). 
Then, it computes the score for each of the nodes (lines~16--22).
The model (described in~\ref{subsec:probing}) is used for the profiling of nodes (line~18). 
The scores are computed by normalizing the predictions and adding the demanded weights (line~20).
Every $x$ seconds the rescheduling phase is triggered for the set of all running tasks.
If the re-execution of the best fit decides on a different target node, the task is migrated to the new host and removed from the current one (lines~9--10).
We show in our evaluation that~$x$, for our specific workload and cluster settings, has minimal impacts on the runtime or the energy efficiency of \sys. We will study this further in future work.

% !TEX root = paper.tex
\section{Architecture}
\label{sec:architecture}
\begin{figure}[!t]
	\centering
	\includegraphics[scale=0.7]{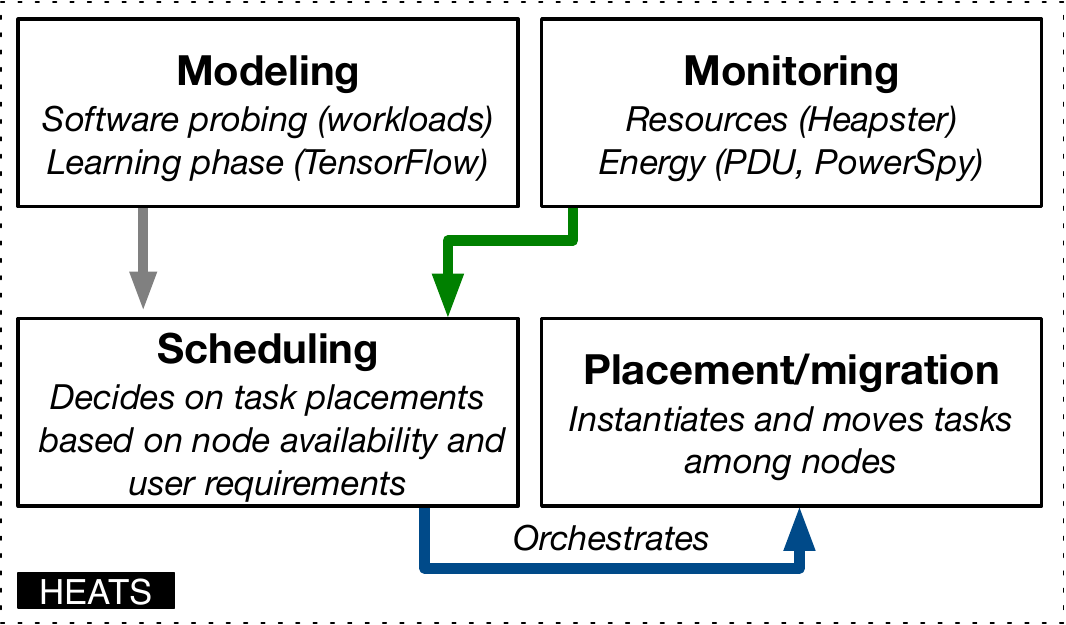}
	\caption{\sys's abstract components and interaction.}
	\label{fig:arch:components}
	\vspace{-15pt}
\end{figure}

\vspace{-2pt}
The architecture of \sys is composed of several interacting components.
Figure~\ref{fig:arch:components} depicts these interactions. 
We describe each of them in details in the remainder of this~section.

%\vs{we need a drawing of the various components, embedded in the current k8s arch}

%\subsection{Modeling}
\label{subsec:probing}
\textbf{Modeling.} The modeling component executes two main operations, namely probing and learning, descibed below.

\begin{figure}[!t]
	\centering
	\includegraphics[scale=0.7]{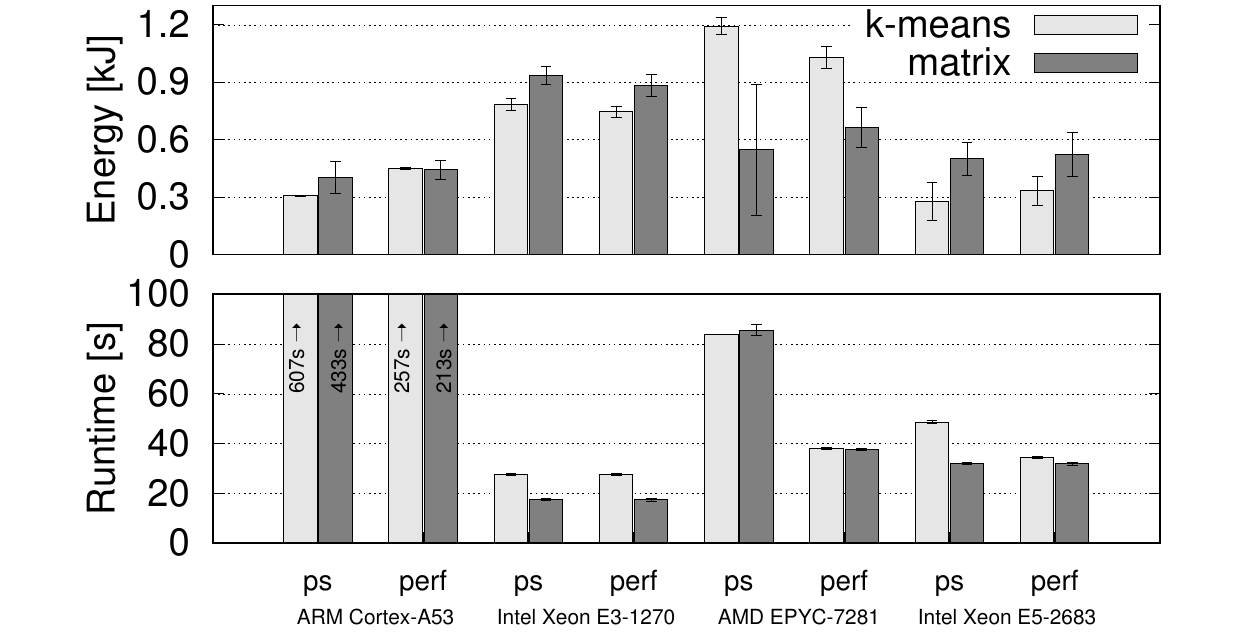}
	\caption{Runtime and energy spent by tasks executing \texttt{k-means} and a matrix multiplication with two different CPU governors: \textit{powersave (ps)} and \textit{performance (perf)}.}
	\label{fig:energy_vs_perf}
    \vspace{-12pt}
\end{figure}

%\textbf{Probing}. 
The probing phase discovers the properties and capabilities of the cluster, \ie, the machines composing it. 
%My idea to sell this is that this is as automatic as possible as a process for users willing to try this scheduler, by making few assumptions (e.g. presence of hardware probes, or software probes).
This probing phase is executed upon the initial setup of \sys, as well as for every major hardware reconfiguration (such as the integration of new machine types in the cluster pool).
%This probing would be part of the initial ‘setup’ of the heat-scheduler (minus some periodic re-execution in case of major hardware reconfigurations).
We implemented this probing so that it also takes care of exploring the performance of the nodes by scaling up and down the frequency of the CPUs~\cite{le2010dynamic}.
We report that, in a typical setup, to produce an accurate model of a new machine usually requires a few hours.
Figure~\ref{fig:energy_vs_perf} shows the results of possible characterizations that this phase can produce, when applied to the machines of our cluster.
In particular, it outputs the runtime and energy requirements of two different families of \emph{probing tasks}.
The energy requirements reported here do not consider the idle state of the machines but of the task itself only.
In this way we can better understand the tasks energy requirements for the differnt types of hardware given.
We show the results with two of such CPU-bound tasks: the aforementioned \textit{k-means} clustering algorithm, as well as a typical matrix multiplication operation.
%\footnote{Our architecture allows for easy integration of additional probing tasks, for instance to capture the behavior under different types of workloads.}
For both types of probing tasks, we observe that the energy requirements can be reduced on a given performance cost for almost every machine type. 
The framework further executes these probing tasks by frequency scaling of the underlying CPUs.
We achieve this by leveraging two different Linux's CPU governors~\cite{linuxcpugov}, \textit{powersave} and \textit{performance}, respectively running the CPU at the minimum and maximum frequency. 
We can observe that, within the same machine type, the energy and performance are largely affected by scaling the CPU frequency. 
The output of this phase is used next. 

%It is also be capable of measuring the machines also via frequency scaling
%to basically build a more refined model (but in first approximation, no scaling is needed)}

%\vs{Questions for Isabelly: how much time it takes to learn the required informations regarding a new machine? Can we provide some numbers here? Orders of few seconds/few minutes?}
%\ir{I cannot tell precisely since we've done it manually and not all at once. But the process evolves running different tasks with different configurations in each of the machines - so it would depend on how many different parameters you would like to account in the probing. But I would say a few hours for a precise initial model.}

%\subsection{Model}
%\label{sec:model}

%\textbf{Learning.} 
The data collected by the probing phase is used to train a multiple linear regression model~\cite{myers1990classical}.
Given a task and its CPU and memory requirements, a fitted regression model is used to predict its energy and performance for each machine type available in the cluster.
We did a preliminary analysis of different machine learning techniques and, for the workload used, TensorFlow\cite{abadi2016tensorflow} presented better results.
We plan a more extensive comparative evaluation of different machine learning algorithms for further work.
%An extensive evaluation of machine learning algorithms with different workloads may lead to a different result, but we consider this extensive evaluation to be orthogonal to the work presented in this paper.
While the probing component constantly records new data, \sys uses it to refine the predictions at a given frequency.
In our evaluation, we execute the learning phase every 24 hours. 
%For the development of the model we rely on TensorFlow \cite{abadi2016tensorflow}, an open-source machine-learning framework.

%\subsection{Monitoring}
\textbf{Monitoring.} Kubernetes is equipped with several tools to monitor resources:
\textsc{cAdvisor}~\cite{cadvisor} has been partially integrated into Kubernetes' node agent \textsc{kubelet}~\cite{kubelet}, and it is capable of measuring resources used by containers.
\textsc{Heapster}~\cite{heapster} exploits the measurements from \textsc{cAdvisor}, aggregates them
and provides means to analyze and monitor the state of the Kubernetes cluster using Grafana~\cite{grafana}. 
Furthermore, \textsc{Heapster} allows us to store the aggregated data in \textsc{InfluxDB}~\cite{influxdb}, a time-series database that supports SQL-like queries to retrieve historical resource measurements of the Kubernetes cluster.
Future versions of \sys will support \textsc{metrics-server}~\cite{metrics_server}.
%\footnote{\textsc{Heapster} was recently deprecated~\cite{heapster_deprecation} in favour of \textsc{metrics-server}~\cite{metrics_server}. We plan to support it in future version of \sys.}

In order to decide whether a task has to be migrated from one node to a different heterogeneous node, the \sys scheduler has to be able to rely on a fine grained resource monitoring system.
Despite the potential capability to gather resource measurements every \SI{5}{\second}, we found out that \textsc{Heapster} cannot reliably deliver these resource measurements at a fixed rate.
A custom resource measurement system was therefore implemented and installed on the Kubernetes nodes, which queries every second the local Docker instances for up-to-date resources used by the containers.
These resource measurements can then be aggregated and used by the \sys scheduler to provide the needed support for migrating tasks.

%In the monitoring phase we make use of Heapster in order to access node as well as pod metrics such as cpu and memory utilization, reservation and capacity.
%Besides this, Heapster also collects events and other signals generated by the cluster. For persisting the data we have set up InfluxDB.
%Regarding the energy monitoring, we have developed a module which will constantly collect the power measurements of the nodes using the external power meters 
%mentioned earlier and fetch this data into the InfluxDB as an additional metric on the node level. As an additional component, we have set up Grafana having 
%the InfluxDB as data source. This provides us with visualization panels that include graphs, tables, and lists based on queries to the data source.

The monitoring component is responsible for actively gathering information regarding the resources currently being consumed at each node by the tasks in execution.
This information is required by the scheduling component (described below) to know which node has sufficient resources for the pending tasks.
\sys leverages some default software probes from \textsc{Heapster} to continuously fetch the hardware resources available on any given node.

Additionally, to access in real-time the current power and energy levels of a node, we assume the availability of hardware monitors that are remotely accessible.
We experimented with two different types of energy monitors, one for server-grade machines and one for low-energy profiles (see Section~\ref{sec:eval}).

%\subsection{Scheduling}
\textbf{Scheduling.} Finally, the scheduling component is in charge of orchestrating the inputs received by the modeling and monitoring components. 
To that end, it first ensures that a prediction for the resources used by the task on the different set of machines is completed.
Then, it combines this prediction with the energy and performance trade-offs, as defined by the end-user, to decide on the best fitting node.
Periodically, the scheduling component reconsiders its past decisions: when a better fitting node is found, a migration decision is taken and the corresponding task is moved to the target node.

%When a given task has to be scheduled the first step is to predict the characteristics it would have in the different machine types by using the monitoring module.
%Afterwards the scheduler uses these estimations combined with the energy and perfomance weights defined by the user to decide the best fitting node.
%Every X seconds this decision is recomputed and if there's a new best fitting node the task is migrated to it.

% !TEX root = paper.tex
\begin{table}[!t]
  \centering
  \setlength{\aboverulesep}{0pt}
  \setlength{\belowrulesep}{0pt}
  \setlength{\extrarowheight}{.74ex}
  \setlength\tabcolsep{1.5pt}
  \rowcolors{1}{gray!10}{gray!0}
  \caption{Hardware characteristics of our cluster.}
  \begin{tabular}{>{\kern-\tabcolsep}lllllllllllll<{\kern-\tabcolsep}}
    \toprule 
	\rowcolor{gray!25}
     & \textbf{Arch.} & \textbf{Cores} & \textbf{Frequency} & \textbf{TDP} & \textbf{Mem.} \\
    \midrule
    ARM Cortex-A53      & \textsc{big.LITTLE} & 4  & \SI{1.4}{\GHz} & 5 \si{\watt} & \SI{1}{\gibi\byte} \\
    AMD Epyc 7281       & amd64 			  & 32 & \SI{2.1}{\GHz} & 155 \si{\watt} & \SI{64}{\gibi\byte} \\
    Intel Xeon E3-1270 v6	& x86 				  & 4  & \SI{3.8}{\GHz} & 72 \si{\watt} & \SI{64}{\gibi\byte} \\
    Intel Xeon E5-2683 v4	& x86 				  & 32 & \SI{2.1}{\GHz} & 120 \si{\watt} & \SI{128}{\gibi\byte}\\
    \bottomrule
  \end{tabular}
  \label{table:hwspecs}
  \vspace{-15pt}
\end{table}

\section{Implementation}
\label{sec:implementation}
\vspace{-2pt}
We base our implementation on Kubernetes~(v1.8), itself implemented in Go~\cite{k8s}.
Custom schedulers can however be implemented in any programming language and connected to the main orchestrator engine via the Kubernetes Scheduler~API~\cite{k8s_scheduler_api}.
\sys is implemented in Python~(v3.6.3) and leverages the Kubernetes Python Client~\cite{k8s_client_python}, a client library for the Kubernetes API~\cite{k8s_api}.
The modeling component (Section~\ref{subsec:probing}) leverages the Python bindings for TensorFlow~(v1.11). 
\sys~is released as open-source and is readily available at~\url{https://github.com/legato-project/heats-scheduler}. %\footnote{Omitted due to double-blind policy. Contact the PC chairs to get an anonymized copy.} % 

% !TEX root = paper.tex
\begin{figure}[!t]
       \centering
       \includegraphics[scale=0.7]{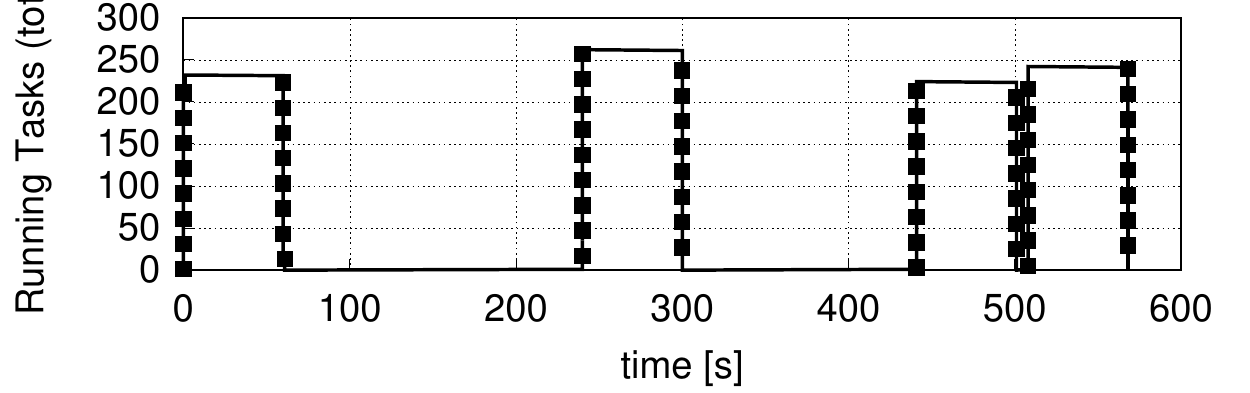}
       \caption{Workload injected by the synthetic trace: tasks arrive in 4 bursts of up to 262 concurrently running containers.}
       \label{fig:distribution}
\end{figure}
\section{Evaluation}
\label{sec:eval}
\vspace{-2pt}
This section presents the experimental evaluation of our \sys prototype.
We first describe the experimental settings.
Then, we describe the synthetic trace used to compare \sys against the default k8s settings.
We compare both schedulers in terms of energy and resource utilization. %,This comparison takes place in the remainder of the section:
%we compare the resources utilization and energy requirements of both schedulers;
We analyse how the user demands (energy/performance ratios) affect the observed performanes.
Finally, we look at the impact of the rescheduling frequency on the overall job~runtime.
%we study how the energy/performance ratio demanded by the users affects the observed performances of the cluster; and finally
%we look at the impact of the rescheduling frequency on the overall job~runtime.
%we further evaluate the \sys scheduler with respect to the impacts of the energy/perfomance ratio
%and rescheduling window used, respectively.

%\subsection{Evaluation settings}
\label{sec:evaluation_settings}
\textbf{Evaluation settings.}
We deploy and conduct our experiments over a cluster composed of 4 different types of machines (see Table~\ref{table:hwspecs}). 
%(1) 4-core ARM Cortex-A53, 1.4 Ghz with 1 GB of RAM and a 5W TPD, (2) a 32-cores AMD Epyc 7281, 2.1 GHz, 64GB of RAM and 155W TPD, (3) an Intel Xeon  
%Table~\ref{table:hwspecs} summarizes their specifications.
Our cluster is composed of 9 machines, where one is the Kubernetes master, orchestrating the deployments and the remaining nodes are workers executing the tasks. 
The 8 worker nodes consist of one AMD, 3 Intel and 4 ARM machines.
The energy consumption is measured using a LINDY iPower Control 2x6M power distribution unit (PDU) for the server type machines and PowerSpy~\cite{banerjee2005powerspy} devices for the three Raspberry Pi. 
The PDU records up-to-date measurements for the active power at a resolution of \SI{1}{\watt} and with a precision of 1.5\%.
We query it up to every second via HTTP.

\begin{figure}[!t]
	\centering
	\includegraphics[scale=0.7]{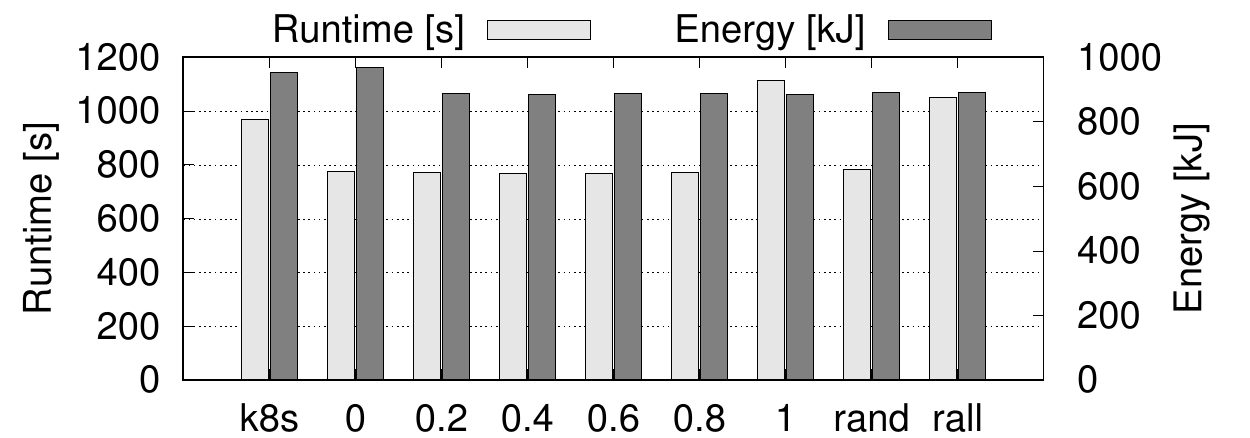}
	\caption{Energy efficiency and impact on the overall runtime of the trace for several scheduling policies.}
	\label{fig:cluster_energy}
	\vspace{-10pt}
\end{figure}

\begin{figure*}[!t]
	\centering
	\includegraphics[scale=0.7]{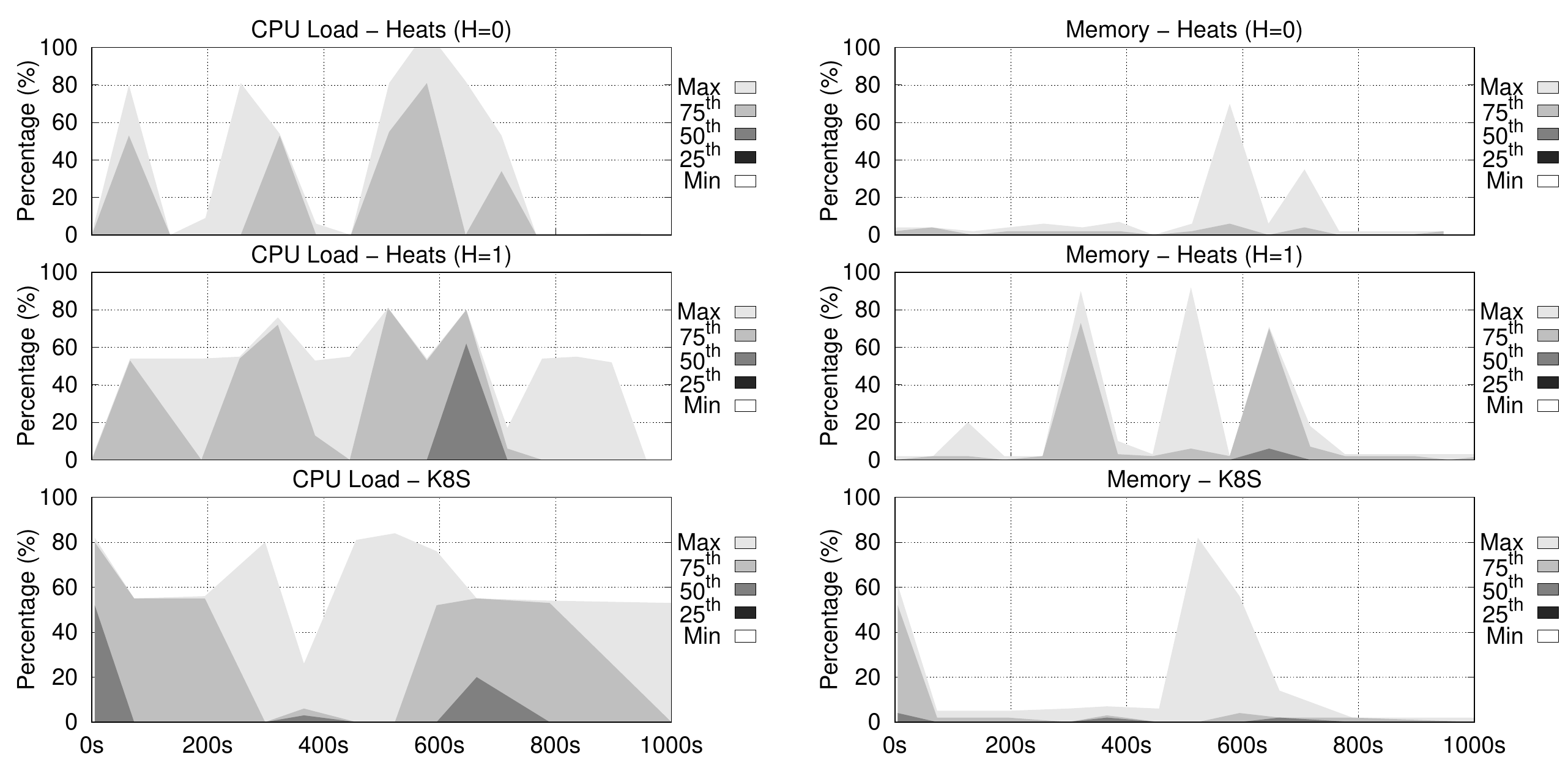}
	\caption{CPU and memory usage distribution (percentiles) across all machines in the cluster. We show these metrics with three different schedulers: \sys in two configurations (H=0, H=1) on the first and second row, Kubernetes in the third row.}
	\label{fig:cluster_cpu_mem}
\end{figure*}
%\subsection{Synthetic trace}
%\label{sec:synthetic_trace}
\textbf{Synthetic trace.} We use a synthetic trace to evaluate the gains and trade-offs of our system.
Figure~\ref{fig:distribution} shows the workload injected by this trace.
We use it to deploy multithreaded tasks executing an iterative implementation of the \texttt{k-means} algorithm in the C programming language.
The program, shipped as statically linked binary for Alpine Linux~\cite{alpine}, executes over a predefined dataset of \num{65536} data points along \num{32} dimensions. 
Once deployed, the tasks will compute clusters by splitting the dataset into blocks processed by two worker threads for a specified maximum number of iterations, chosen randomly in the range of \numrange{500}{1000}.
The result is stored as file inside the container's image. 
In total, \num{480} \texttt{k-means} jobs are deployed following four bursts over \num{10} minutes, executed randomly within a timeframe of \num{150} seconds.
The same sequence of pseudo-random numbers is ensured upon every run of a trace by using a fixed random seed. %to the program running the trace.
%\cg{Add sentences explaining the injected distribution}

%\subsection{Kubernetes vs. \sys}
%\label{sec:kube_vs_heater_scheduler}
\textbf{Kubernetes vs. \sys.} First, we compare the CPU load induced on the cluster by \sys against the default scheduling policy of Kubernetes.
Figure~\ref{fig:cluster_cpu_mem} shows these results.  %of the execution of the trace on our cluster.
We observe how the load patterns are very similar and closely follow the arrival pattern of the tasks in the trace  Figure~\ref{fig:distribution}).
We conclude that the \sys scheduler does not deteriorate the lifetime of the processors by artificially stressing them. 
Next, we look at the memory usage across the cluster. 
Figure~\ref{fig:cluster_cpu_mem} shows this for two different \sys configurations, one for performance ($e$\,=\,0, $p$\,=\,1), the other for energy-efficiency ($e$\,=\,1,$p$\,=\,0).
The memory load of two schedulers induce similar patterns.
%We compare the resources utilization and energy requirements of the cluster while running the trace describe in \ref{sec:synthetic_trace}.
%Figure~\ref{fig:cluster_cpu} and Figure~\ref{fig:cluster_memory} show the cpu and memory utilization, respectively.
%\begin{figure}[!t]
%	\centering
%	\includegraphics[scale=0.7]{plots/heats-burst/memory.pdf}
%	\caption{Memory usage averaged across all machines in the cluster.}
%	\label{fig:cluster_memory}
%\end{figure}
We compare the energy efficiency of the default Kubernetes scheduler against \sys.
Figure~\ref{fig:cluster_energy} presents the total runtime and the cumulative energy consumption of the cluster throughout the execution of the trace, 
including their idle state requirements.
We compare five different approaches:
(1)~the default scheduler (\texttt{k8s}),
(2)~\sys configured to deploy tasks on the fastest possible machines, ignoring any energy concerns (\texttt{e0p1}),
(3)~\sys trying to be as energy-efficient as possible.
Moreover, for the sake of comparison, we include the results achieved by
(4)~a fixed \emph{H} value chosen out of our practical experience (\texttt{rand}, for $e$\,=\,0.618, $p$\,=\,0.382) and (5) other variations of the H-value.
When compared to the default scheduler (\texttt{k8s}), (2) performs better on an energy cost of 1.5\% while (3) performs worse but presents 7.1\% of energy savings. 
Besides, when compared to each other, (2) performs better while (3) is more energy efficient. 
Finally, for approach (4) we can observe that the runtime as well as the energy consumption are in between the observations for approach (2) and (3). 
Therefore, we can conclude that our observations follow the expected behaviour.

%Figure~\ref{fig:cluster_energy} shows the cummulative energy consumption of the cluster for the execution of the entire trace while scheduling the tasks with the default Kubernetes stratary and the heater scheduler configure to focues either entirely on energy efficiency or performance.

%\subsection{Energy vs. performance weights}
%\label{sec:energy_vs_performance_weights}
\textbf{Energy vs. performance weights.} The value chosen for the \emph{H} parameter is of paramount importance, especially when considering the resulting energy costs and impact on the overall runtime of the jobs.
To better understand this aspect, we choose 6 different configurations (from 0 to 1, by increments of 0.2), for different energy/performance ratios, 0 being the least and 1 the most energy-efficient versions.  
We compare the achieved results with a \sys configuration that randomly select the value of \emph{H}, mimicking a customer with no particular requirements.
%We evaluate the thrad-off between energy and performance by varying the energy/performance ratio of the \sys scheduler.
Figure~\ref{fig:cluster_energy} shows our results.
For each configuration, we show the cumulative energy costs (in~kJ) and the achieved runtime, respectively on the left and right vertical bars.
We observe how the configurations achieve similar results, with a sensible deviation only with the less energy efficient variant.
While these results require further investigations, we believe them to be of practical interest for end-users. 
We intend to confirm these by evaluating the same configurations on real-world traces, where the variations of the \emph{H} parameter might have more impact.

\section{Related Work}
\label{sec:rw}
There is a large body of literature on scheduling, deployment and migration policies, mainly driven by the strong momentum on green computing~\cite{hooper2008green}.
Here we focus on energy-related scheduling policies in a container/task-based deployment setting, leaving out research on optimization problems specifically geared toward reducing energy costs~\cite{ll-es-greedy}.

% Wang, 2014, Energy-aware Data Allocation and Task...
Wang \etal~\cite{rw:rts} proposes two polynomial-time algorithms, one for energy-aware heterogeneous data allocation and another for task ratio greedy algorithm. 
The algorithms schedule real-time constrainted tasks of applications on heterogeneous multiprocessor systems using integer linear programming. %to meet time constraints.
Simulations compare these two algorithms against a greedy algorithm on two heterogeneous multiprocessor systems. 
\sys does not support real-time constrains. 
On the other hand, we fully implemented \sys and perform deployments of real code as well leveraging software and hardware monitors to gather energy-related metrics.
% are targeted and real metrics are measured without the use of simulation. 
%Finally, our testbed consists of four heterogeneous types of machines, providing more insights on the possible energy~savings.
%\cg{compare Wang \etal energy savings to our values once available}

GenPack~\cite{havet2017genpack} proposes an energy-saving mechanism inspired by the JVM's garbage-collectors. 
It migrates containers from young (unstable) to old (stable) generations of machines. 
Its architecture, based on Docker Swarm, is similar to the one built for \sys.
However, GenPack ignores the user-demanded trade-offs of the jobs, and containers are migrated across the cluster only by observing the \emph{stability} of the jobs.

Partial Optimal Slacking (POS)~\cite{ees} is an energy-efficient scheduling approach based on the concept of \emph{task slacking} with the objective to lower the processing speed of a processor executing a task without affecting other tasks.
POS achieves this by using DVFS techniques~\cite{shekar2010energy}. 
The frequency scaling support in \sys is currently exploited during the modeling phase. 
We intend to further leverage this feature, for instance while migrating
tasks for isolation within the processor, in case two unlike tasks
are running on the same core but require different frequency and voltage levels.

%Power and performance management for parallel computations~\cite{ll-es-greedy}

% !TEX root = paper.tex
\section{Conclusion and Future Work}
\label{sec:conclusion}
\vspace{-2pt}
We presented \sys, a novel task-based scheduling system for heterogeneous clusters. 
\sys learns about the properties of the machines in the cluster, schedules and possibly migrates tasks to the best-fitting node currently available.
Our experimental evaluation reveals that \sys can yield considerable energy savings depending on the type of resources at hand, the workload and the desired energy/performance ratio.

We envision to extend this work along the following directions.
First, we will explore how per-core pinning and per-core frequency scaling can further improve the achievable energy savings.
This will have side-effects in the learning (\ie, probing) phase, which will need to be extended.
Second, we will extend the design and implementation of our prototype to account for migrations between heterogeneous devices, \emph{i.e.}, from an x86 processor onto a GPU.
This will require producing binaries targeting different architectures and managed by the same scheduler.
Finally, we will evaluate \sys against real-world traces, \emph{e.g.}, Borg~\cite{clusterdata:Wilkes2011} and Azure~\cite{cortez2017resource}.
\section*{Acknowledgments}
\vspace{-2pt}
	The research leading to these results has received funding from the European Union's Horizon 2020 research and innovation programme under the LEGaTO Project (\href{https://legato-project.eu/}{legato-project.eu}), grant agreement No~780681.

{
\footnotesize
\printbibliography
}

\end{document}